\documentclass[aps,prl,reprint]{revtex4-1}
\usepackage{blindtext}

\usepackage[normalem]{ulem}
\usepackage{amsmath}
\usepackage{amssymb}
\usepackage{color}
\usepackage{graphicx}
\usepackage{lineno}
\usepackage{subfigure}

\begin{document}

\title{Scalable Discovery of Fundamental Physical Laws: Learning Magnetohydrodynamics from 3D  Turbulence Data}
\author{Matthew Golden}
\email{matthew.golden@gatech.edu}
\author{Kaushik Satapathy}
\author{Dimitrios Psaltis}
\affiliation{School of Physics, Georgia Institute of Technology}

\begin{abstract}
The discovery of dynamical models from data represents a crucial step in advancing our understanding of physical systems. Library-based sparse regression has emerged as a powerful method for inferring governing equations directly from spatiotemporal data, but current model-agnostic implementations remain computationally expensive, limiting their applicability to data that lack substantial complexity. To overcome these challenges, we introduce a scalable framework that enables efficient discovery of complex dynamical models across a wide range of applications. We demonstrate the capabilities of our approach, by ``discovering'' the equations of magnetohydrodynamics (MHD) from synthetic data generated by high-resolution simulations of turbulent MHD flows with viscous and Ohmic dissipation. Using a library of candidate terms that is $\gtrsim 10$ times larger than those in previous studies, we accurately recover the full set of MHD equations, including the subtle dissipative terms that are critical to the dynamics of the system. Our results establish sparse regression as a practical tool for uncovering fundamental physical laws from complex, high-dimensional data without assumptions on the underlying symmetry or the form of any governing equation.
\end{abstract}

\maketitle

Discoveries in physics often emerge through a well-charted trajectory. This typically begins with classifying observations into distinct categories, progresses to uncovering empirical algebraic relationships that enable forecasting but lack universality, advances to the development of dynamical models where these empirical laws emerge as solutions to differential equations, and culminates in the abstraction of these models through principles of symmetry and invariance. Modern machine learning (ML) tools based on different neural-network architectures have proven highly effective in advancing the first two steps of scientific discovery~\cite{mehta2019,Smith2023}: classifying phenomena and uncovering empirical algebraic relationships. However, their utility in the third step, i.e., the development of generalizable dynamical models, remains limited. 

One fundamental reason for this is the fact that the systems we now study are far more complex than the simpler scenarios of earlier discoveries, making it nearly impossible to derive simple empirical models directly from the data. Modern datasets often contain layers of emerging complexity, obscuring the simplicity of the underlying models. This underscores a fundamental challenge: when the data are complex and chaotic but the underlying models are simple, can we bypass the intermediate step of uncovering empirical relationships and derive dynamical models directly from the data using modern computational tools? As an example, could we feed an ML algorithm with experimental data from a turbulent flow in the presence of magnetic fields and have it discover the equations of magnetohydrodynamics? 

Library-based sparse regression has rapidly emerged as an effective method for inferring dynamical equations directly from  data~\cite{crutchfield1987,brunton2016discovering,rudy2017data}. The goal of this approach is to find the functional forms of a set of differential operators ${\cal F}$ such that the corresponding dynamical (and generally non-linear) equations
\begin{equation} 
    {\cal F}\left(u_i,\frac{\partial u_i}{\partial x_j},\frac{\partial^2 u_i}{\partial x_j^2},...;x_j\right)=0\;,\quad
    i=1,...,N; j=1,...,M
    \label{eq:dyn_system}
\end{equation}
best describe a particular set of data for $N$ physical quantities in $M$ dimensions, including time. Naturally, there could be an infinite number of possible such equations. To address these degeneracies, the sparse regression approach identifies the dynamical model with the minimal number of terms necessary to describe the data. This sparsity-promoting methodology faces no difficulty in learning partial differential equations from spatiotemporal data \cite{rudy2017data, gurevich2019robust} and has shown success when applied, e.g., to experimental measurements in quasi-2D fluids~\cite{reinbold2021robust}, active nematic turbulence~\cite{golden2023physically,joshi2022data,robertson2023continuum}, and cell mechanics from protein images~\cite{schmitt2024machine}.

\begin{figure}[t]
    \includegraphics[width=0.9\linewidth]{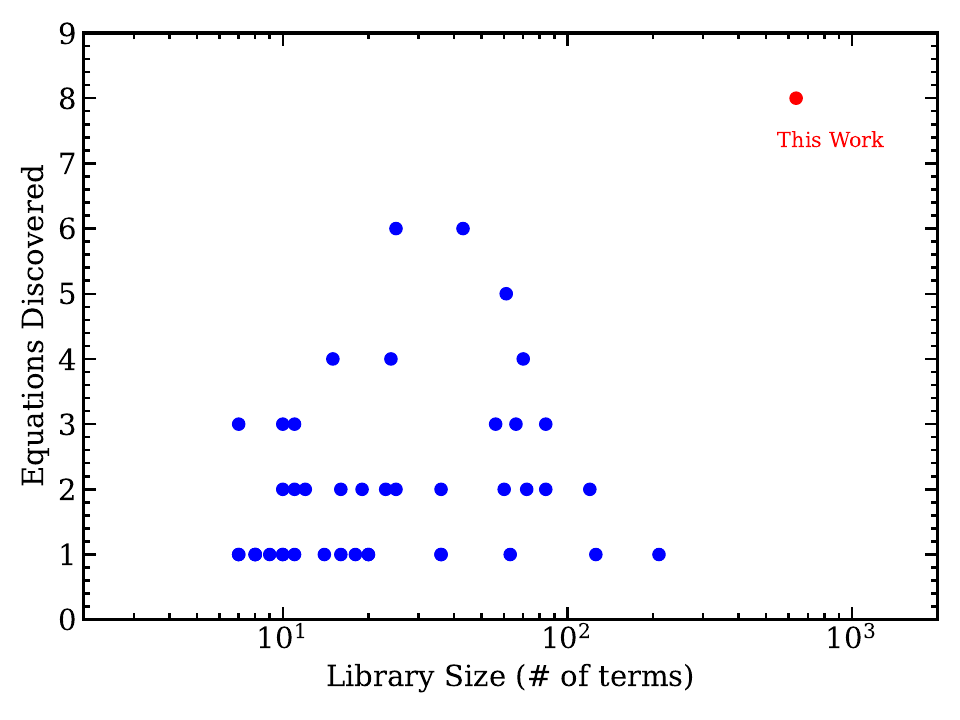}
    \vspace*{-0.5cm}
    \caption{Number of equations discovered plotted against the size of the library of potential terms. Blue points represent the capabilities of a large selection of previous efforts, while the red point shows the significant advancements made possible by the method introduced in this Letter. 
    \label{fig:prior}}
    \vspace*{-0.5cm}
\end{figure}

Despite its promise, model discovery using sparse regression has found limited applicability to relatively simple problems with the algorithms searching through a limited range of potential dynamical terms (see Fig.~\ref{fig:prior}). The reason is that current implementations are either computationally expensive with poor scaling, lack generality, require assumptions about the discovered equations, or are insensitive to small but important coefficients. Expanding this approach to enable the versatile discovery of complex dynamical models in various settings requires significant advances. In this {\em Letter\/}, we introduce a scalable framework for data-driven model discovery, addressing critical issues and providing solutions that enable wide and efficient discovery of dynamical models across a broad range of physics applications. 

To demonstrate the power of our approach, we focus on the challenging problem of ``learning'' the equations of magnetohydrodynamics (MHD) using data from numerical simulations of freely decaying turbulent MHD flows, incorporating viscosity and Ohmic dissipation. From a library of candidate terms that is $\gtrsim 10$ times larger than typical previous studies, we accurately recover the full set of MHD equations, including the subtle dissipative terms critical to the dynamics of the system. Our results demonstrate that sparse regression has now matured to enable the discovery of physical models based on data that exhibit intricate, seemingly chaotic patterns without \textit{a priori} knowledge of physical symmetries.

We focus on ``learning'' MHD equations for several compelling reasons. First, early studies demonstrated poor performance in discovering MHD equations from simulation data in simple, non-turbulent configurations. Such nearly laminar setups lack the diversity of realizations needed to uncover the full set of terms in equations~\cite{vasey2023}. This highlights that {\em leveraging the complexity of turbulent flows to discover dynamical equations is not only inescapable in some settings but is also often necessary.\/} Second, MHD equations constitute a system of four coupled partial differential equations with significant complexity, involving one scalar quantity (density) and two vector quantities (velocity and magnetic field), along with numerous cross terms. Accounting for the individual components, this requires our algorithm to identify 8 separate differential equations from the data, which is beyond the vast majority of prior model discovery studies. Third, by incorporating Ohmic dissipation and viscosity at large Reynolds and magnetic Reynolds numbers, we challenge the method further, since it needs to pay attention to both the larger advective scales and the smaller dissipative ones. Finally, the divergence-free condition for the magnetic field introduces an elliptical constraint with only spatial derivatives rather than a dynamical equation with time derivatives. This adds an additional layer of complexity that earlier sparse regression algorithms were not equipped to handle. 

\begin{figure}[t]
    \centering
    \includegraphics[width=0.4\textwidth,trim={1cm 1cm 0cm 3.5cm},clip]{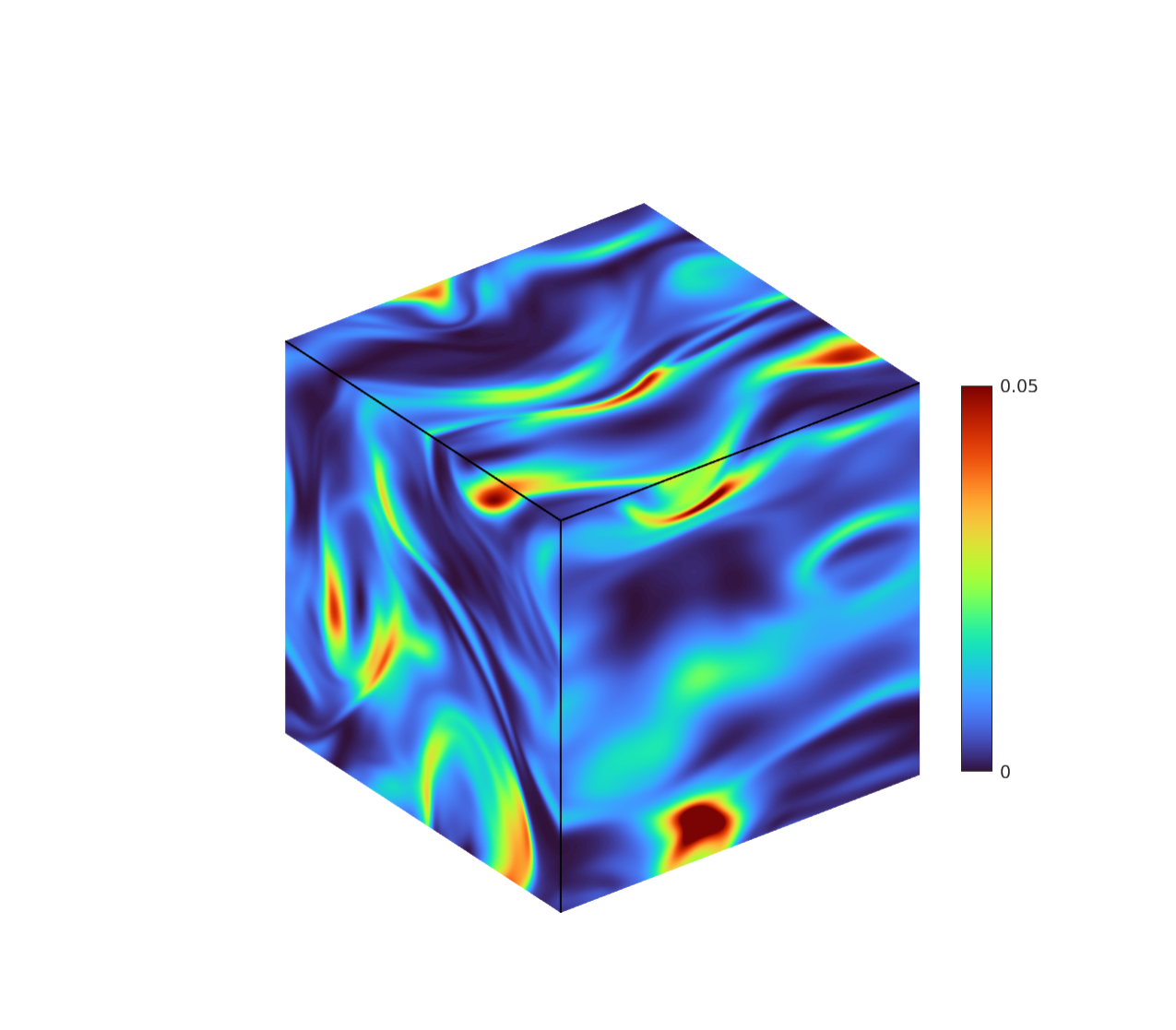}
    \vspace*{-0.5cm}
    \caption{Snapshot of kinetic energy density $\rho {\bf u} \cdot {\bf u}$ in our simulation of fully developed 3D MHD turbulence. }
    \label{fig:simulation}
    \vspace*{-0.5cm}
\end{figure}

To generate the numerical data, we use the \texttt{PENCIL} code \cite{brandenburg2020pencil} and solve the MHD equations on a periodic 3D spatial domain. The governing equations are
\begin{eqnarray}
    \nabla \cdot {\bf B} &=& 0, \label{eq:Gauss}\\
    \partial_t \rho + \nabla \cdot ( \rho {\bf u}) &=& 0,\label{eq:continuity}\\
    \partial_t {\bf B} + \nabla \times ( {\bf u} \times {\bf B}) &=& \eta \nabla^2 {\bf B}, \label{eq:induction}\\
     \partial_t( \rho {\bf u}) + \nabla \cdot (\rho {\bf u}  {\bf u}) &=& \nabla \cdot ( {\bf BB} ) - \nabla \left( \rho + \frac12  {\bf B} \cdot {\bf B} \right) \nonumber\\
    && \qquad\qquad+ \nu \rho \nabla^2 {\bf u},\label{eq:momentum}
\end{eqnarray}
where we have set $\nu = \eta = 4\times 10^{-4}$; these correspond to Reynolds and magnetic Reynolds number of 2500. We use a mean magnetic field $\langle B_y \rangle = 0.1$ to prevent magnetic energy from dissipating fully. We initially drive the simulation by long wavelength Kolmogorov forcing until we observe a satisfactory inertial range. We then turn off forcing and allow the turbulence to freely decay. This transient decaying turbulence is the focus of this work. Figure \ref{fig:simulation} shows a snapshot of the resulting kinetic energy density. Before utilizing these data, we first introduce three major advances in the sparse regression method.

\noindent {\em Library generation.---\/} The first advance is constructing a library of potential terms that could contribute to the various equations. Hereafter, to ensure lack of ambiguity in distinguishing multiplicative and additive terms in the equations, we use terminology developed to describe written language. In particular, we define an ``alphabet'' of potential terms (e.g., simple terms that depend on only one dependent variable such as $u,x,\sin x, e^u, \partial u/\partial x, \partial^2u/\partial x^2, ...$) from which we generate all possible ``words'' ($w_k$),  i.e., products of terms such as $u\cdot u \sin (x) \partial^2u/\partial x^2$, and all possible ``sentences'', i.e., linear combinations of words with constant coefficients ($c_k$). The objective of the method is then to identify the null sentences, i.e., relations of the form
\begin{equation}
    \sum c_k w_k=0\;.
    \label{eq:sentence}
\end{equation}
In this expression, the words $w_k$, which we also refer to as ``features'', are calculated directly from the data, whereas the coefficients $c_k$ need to be inferred by the algorithm. 

It is clear that, as the size of the alphabet increases, the number of potential words that the algorithm needs to explore increases (nearly) exponentially~\cite{golden2024scalable}. Consequently, most libraries used for model discovery thus far have been relatively small (see Fig.~\ref{fig:prior}). As applications to higher-dimensional data have become desirable, symmetry-covariant libraries have instead been constructed that explicitly respect known continuous symmetries to drastically reduce their sizes~\cite{gurevich2019robust, golden2023physically}. However, {\em some of the most intriguing physical phenomena involve symmetry-breaking terms,\/} which these approaches cannot accommodate. In this work, we avoid symmetry assumptions altogether and demonstrate that large-library sparse regression is both feasible and effective for uncovering complex spatiotemporal dynamics.

To this end, we construct a library that includes all MHD fields and their first-order spatio-temporal derivatives. It was crucial to the success of sparse regression that density was split into a constant mean and fluctuations $\rho = \rho_{\textrm{mean}} + \tilde{\rho}$. The resulting 35 letter alphabet is
\begin{align}
\mathcal{A} = \{ \tilde{\rho}, u_x, u_y, u_z, B_x, B_y, B_z, \partial_t \tilde{\rho}, \partial_x \tilde{\rho}, \cdots\}\;.
\end{align}
We then consider all one- and two-letter words. We further augment the library with the effective three-letter words needed to recover momentum transport in Equation \eqref{eq:momentum} to produce a library with 627 words. Including all three-letter words would further increase the size of the library by an order of magnitude. This library is sufficiently rich to uncover all governing MHD equations. 

\noindent {\em Weak Formulation.---\/} The second advance involves the evaluation of the words (or features) $w_k$ using simulation data. The weak formulation, which replaces pointwise evaluation of the differential terms in the library with volume-weighted integrals, has been a critical step forward in this step~\cite{gurevich2019robust, messenger2021weak, messenger2024weak, golden2023physically}. In this formulation, we do not look for equations of the form~(\ref{eq:sentence}), but rather for integrals of this equation within small spatio-temporal volumes $\Omega_l$ in the domain of solution, i.e.,
\begin{equation}
   \sum_k c_k  \frac{1}{\Omega_l S_k}\int_{\Omega_l} \phi w_k dV = 0 \Rightarrow {\bf G} {\bf c}=0\;.
    \label{eq:weak}
\end{equation}
In the last equality, we have defined the vector of coefficients ${\bf c}\equiv \{c_k\}$ and the feature matrix
\begin{equation}
    {\bf G}=\{ G_{kl}\}\equiv \frac{1}{\Omega_l S_k}\int_{\Omega_l} \phi w_k dV\;,
\end{equation}
with each component $G_{kl}$ measuring the volume average of word $w_k$ over the spatiotemporal volume $\Omega_l$ and characteristic magnitude scale $S_k$. We choose these volumes randomly to cover the data domain; $\phi$ is an analytic window function that drops smoothly to zero towards the edges of each small volume. The weak formulation helps avoid evaluating numerical derivatives from noisy data by using integration by parts. Moreover, the integrals generate averages over spatiotemporal volumes and hence decrease the influence of noise, while the window function masks discontinuities in the underlying fields~\cite{golden2023physically}. Evaluating these integrals over the sample spatiotemporal volumes requires more than 10 TFLOP (tera floating
point operations). To achieve this within practical computational times, we implemented advanced optimizations in parallelization and memory access, significantly enhancing the efficiency and scalability of our algorithm.

\begin{figure}[t]
    \centering
    \includegraphics[width=0.4\textwidth]{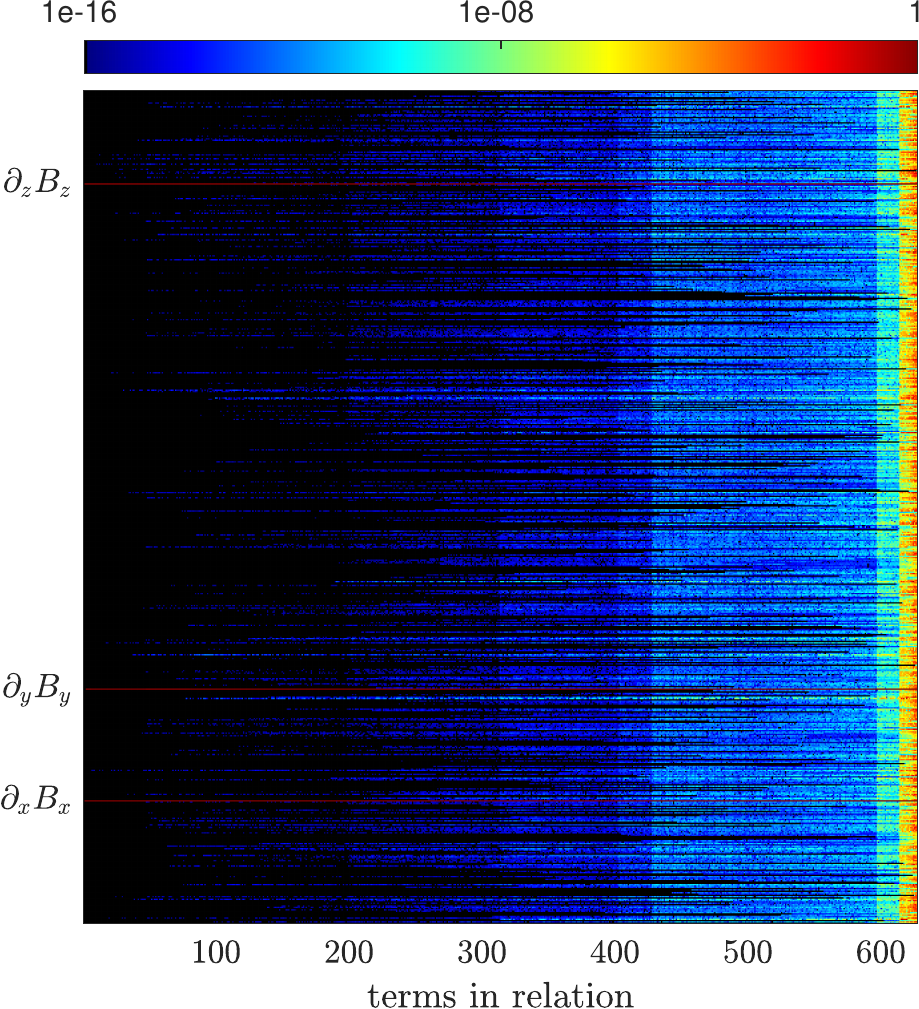}
    \vspace*{-0.1cm}
    \caption{A typical evolution of the greedy sparse regression procedure that identified the equation $\nabla \cdot {\bf B} = 0$. The colorbar shows, for each iteration, the contribution to the residual of each term normalized by the maximal contribution; removed terms correspond to black pixels. At the onset, all terms are considered and at each iteration (from right to left), the term with smallest contribution to the residual is being greedily removed. The ordering of library terms is arbitrary.
    \label{fig:sparse_reg}}
    \vspace*{-0.5cm}
\end{figure}

\noindent {\em Sparse regression.---\/} The third and key advance is devising a new algorithm that identifies the smallest number of non-zero coefficients $a_k$ that make the left-hand-sides of these equations vanish. Sparse regression usually balances quantitative accuracy with parsimony in one of two ways: through the minimization of a sparsity promoting loss function~\footnote{For example, LASSO~\cite{santosa1986linear}, elastic net~\cite{zou2005regularization}, MIO-SINDy~\cite{bertsimas2023learning}.} or by iteratively trimming terms in the equation that have negligible contributions \footnote{SINDy~\cite{brunton2016discovering}, SINDy-PI~\cite{kaheman2020sindy}, Subspace Pursuit~\cite{dai2009subspace}, SPIDER~\cite{gurevich2019robust}.}. Sparse regression methods can be further split into implicit and explicit algorithms. In the explicit case, one particular term (usually the time derivative of one of the quantities) is assumed to be present in an equation, whereas in the implicit case, all library terms are placed on equal footing. Implicit schemes can, therefore, identify not only dynamical equations but also constraints. This capability is critical for our application, since the MHD equation $\nabla\cdot B=0$ is a non-dynamical constraint. Unfortunately, this capability also makes implicit methods inherently more computationally expensive.  

We have developed an efficient algorithm for implicit sparse regression, SPRINT, which drastically reduces the runtime required to parse large libraries of terms from astronomically high values to computationally feasible levels~\cite{golden2024scalable}. SPRINT estimates the sparse coefficient vector ${\bf c}$ that satisfies equation~(\ref{eq:weak}) by iteratively minimizing the normalized residual $r({\bf c}) \equiv \|{\bf G} {\bf c} \|_2 / \| {\bf c} \|_2$, where the symbol $\| . \|_2$ denotes the 2-norm of a vector. Figure~\ref{fig:sparse_reg} shows a visualization of the algorithm and its greedy removal of potential terms until it identifies a sparse equation.

Our method of sparse regression avoids thresholding by coefficient magnitude altogether. This is critical for the recovery of small but important coefficients, such as the ones describing the dissipation of the velocity and magnetic fields. Moreover, it is a significant improvement over methods that minimize a sparsity promoting cost function or threshold small coefficients, for which one must guess {\em a priori\/} the value of a hyperparameter (i.e., the threshold) that allow these small coefficients survive. 

\begin{figure}[t]
    \centering
    \includegraphics[width=0.4\textwidth]{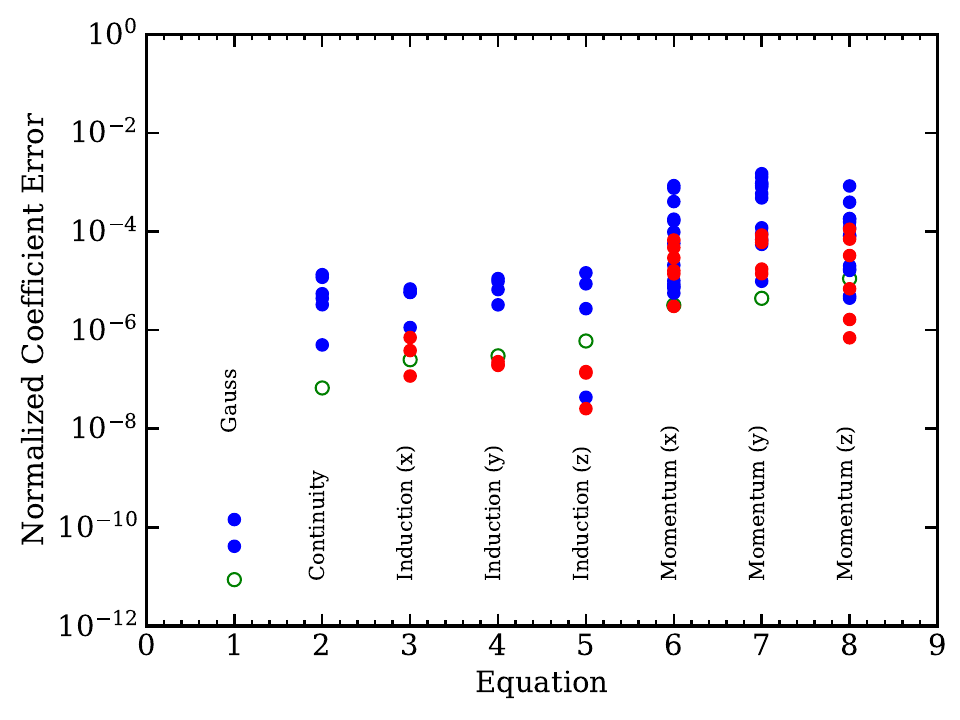}
    \vspace*{-0.1cm}
    \caption{Normalized error in the values of the inferred coefficients of the 8 MHD equations discovered by our algorithm. Red points correspond to dissipative terms while green circles show the residuals. The coefficients in Gauss' law are inferred at machine level, as this constraint equation is not evolved forward in time but is enforced at each time step, therefore not acquiring numerical errors. 
    \label{fig:coefficients}}
    \vspace*{-0.5cm}
\end{figure}

Our algorithm identified 8 differential equations that describe the turbulent MHD flow, with coefficient errors and residuals shown in Figure \ref{fig:coefficients}. The 3-term Gauss's law was enforced exactly in the simulation by evolving the vector potential, so it is found with a residual of $\sim 10^{-11}$ and coefficient errors $\sim 10^{-10}$, which are comparable to machine accuracy. The 7-term continuity equation and the three components of the 8-term induction equation are found next with residuals that reflect the accumulation of numerical errors in the simulations. The three components of the 20-term momentum equation are also found with similar residuals. For the $y$-component of the momentum equation, the algorithm straggles to identify the higher-order dissipative term $\tilde{\rho} \partial_y^2 u_y$, even though it does discover the lower-order one, $\partial_y^2 u_y$. This is due to the mean magnetic field constraining the flow such that the physical magnitude of this term is below the noise floor of our weak-form integrals. 

Our work demonstrates the power of large-library sparse regression to uncover governing equations directly from complex spatiotemporal data. By leveraging the weak formulation and scalable implicit sparse regression techniques, we have demonstrated the feasibility of identifying complex dynamical systems from extensive libraries without relying on symmetry constraints. This approach enables the discovery of fundamental physical laws, even in systems exhibiting intricate, turbulence-driven dynamics.

\bibliographystyle{apsrev4-1} 
\bibliography{biblio.bib} 

\clearpage



   
\section{Supplemental Material}

Here we provide additional details on the generation of the synthetic data, the hyperparameters of the sparse regression algorithm, an estimate of the computational requirements, and our approach to removing degeneracies among the learned equations.

\bigskip

\noindent {\em Synthetic Data Generation.---\/} To generate the synthetic data, we use the \texttt{PENCIL} code \cite{brandenburg2020pencil} and solve the MHD equations on a periodic 3D spatial domain, with $256^3$ grid points, using sixth order differencing in space and third order integration in time. As discussed in the text, we have set the viscosity and magnetic diffusivity coefficients to $\nu = \eta = 4\times 10^{-4}$; these corresponds to Reynolds and magnetic Reynolds number of 2500. We drive the turbulence by adding a Kolmogorov forcing ($f_d$), of the form $f_d = A \cos(k_d z) \hat{x}$ to the right hand side of the momentum equation \eqref{eq:momentum}. We set the amplitude of the forcing to be $A = 0.005$ and achieve a large wavelength driving by setting the driving wavenumber $k_d = 2.0$. We evolve the system for 500 sound crossing times to allow the turbulence to fully develop. We then turn off the forcing and let the turbulence freely decay for 32 sound crossing times, during which we perform model discovery 

\bigskip

\begin{figure}[t]
    \centering
    \includegraphics[height=0.4\textwidth]{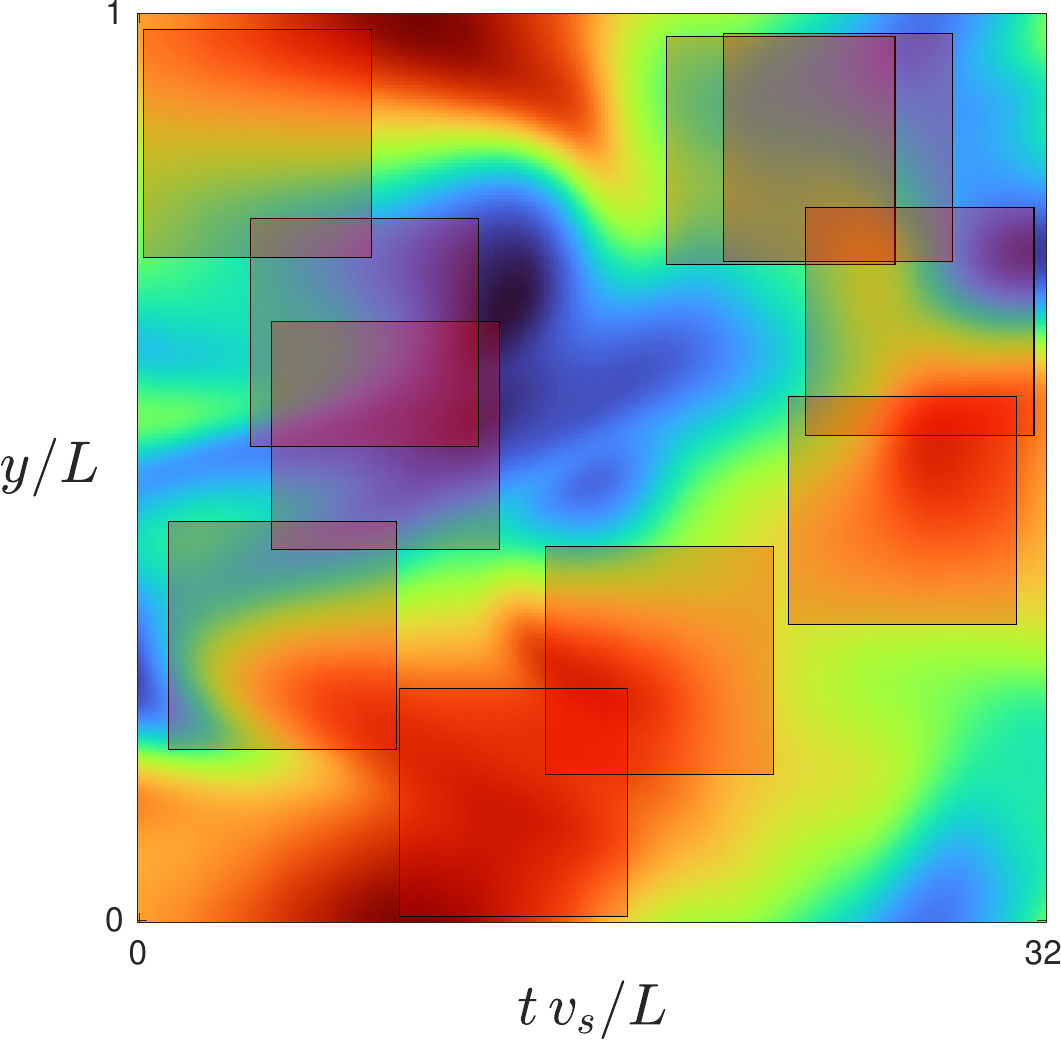}
    \caption{Cross section of a small region of the four-dimensional spacetime, showing some of the spatiotemporal volumes (red squares) used in the weak formulation of the model discovery algorithm. The color shows $B_x(y,t)$ at a particular constant slice in $x$ and $z$. \label{fig:volumes}}
    \vspace*{-0.5cm}
\end{figure}

\noindent {\em Algorithmic Hyperparameters.---\/} In order to sample a large number of realizations of the dynamics, we consider 1376 domains $\Omega_l$ using spatiotemporal volumes of $64^4$ gridpoints each and a polynomial window function $\phi = \prod_{\mu=1}^{4} (1-\tilde{x}_\mu^2)^\beta$, with $\beta = 8$~ \cite{messenger2021weak,gurevich2019robust}. Figure~\ref{fig:volumes} shows a cross section of the 4D spacetime of the simulation, together with some sample spatiotemporal volumes. The benefit of the weak formulation is that derivatives do not need to be evaluated if the term is of the form $\partial^n u$, where $\partial$ is an arbitrary partial derivative. For terms of the form $u \partial v$ with $u\neq v$, integration by parts is not possible. In this case, we estimate numerical derivatives with centered second-order finite differencing to reduce memory costs. 

We take advantage of the Leibniz rule to manipulate the library: the goal is to maximize the number of terms that can be integrated by parts. For example, the span of the terms $\{ u \partial v, \, v \partial u \}$ is the same as $\{ \partial (uv), \, u\partial v \}$. The latter choice of basis is preferable because the weak formulation can be employed for the first term. This does not change the size of our library or the equations we can capture. We apply this consistently by defining a canonical ordering of fields. We simply replace non-canonical terms $v\partial u$ with their Leibniz equivalent $\partial(uv)$. In this study, we consider all one and two-letter words spelled with the symbolic alphabet $\mathcal{A}$.
\begin{align}
\mathcal{A} = \{ 
&\tilde{\rho}, u_x, u_y, u_z, B_x, B_y, B_z, \nonumber \\
&\partial_t \rho, \cdots, \partial_x \tilde{\rho}, \cdots, \partial_y \tilde{\rho}, \cdots, \partial_z \tilde{\rho} , \cdots \},
\end{align}
where $\tilde{\rho}$ is the mean-subtracted density $\tilde{\rho} \equiv \rho - \rho_{\rm mean}$. The alphabet $\mathcal{A}$ contains all field components and first order spatiotemporal gradients. This makes $|\mathcal{A}| = 35$. We augment the full library with second order derivatives, density weighted second order derivatives, and advective terms of the form $\partial_x (\tilde{\rho} u_x u_y)$ expected in the momentum equation. Considering all possible three-letter words would increase the library size by an order of magnitude, so we resort to this augmented, primarily two-letter library as a demonstration of our model discovery tools.

SPRINT estimates the sparse coefficient vector ${\bf c}$ that satisfies equation~(\ref{eq:weak}) by iteratively minimizing the normalized residual $r({\bf c}) \equiv \|{\bf G} {\bf c} \|_2 / \| {\bf c} \|_2$, where the symbol $\| . \|_2$ denotes the 2-norm of a vector. The residual is exactly minimized by a right singular vector of ${\bf G}$ associated with the minimal singular value: $ \min_{\bf c} r({\bf c}) = \sigma_{\textrm{min}}$. We first compute this exact minimum with the singular value decomposition (SVD) and then efficiently compute the impact of each ``word'' on $\sigma_{\textrm{min}}$ by setting each element of ${\bf c}$ to zero. We then remove the ``word'' from the ``sentence'' that causes the smallest increase in residual and repeat. This greedy procedure generates a curve of increasingly sparse models with monotonically increasing residuals. We then select the optimal sparsity by identifying the point at which removing one term introduces a substantial change in the residual, i.e., when between two iterations the residual increases by $r_{i-1} / r_{i} > \gamma$. 

\begin{figure*}[t]
    \centering
    \begin{subfigure}[] {\includegraphics[height=0.25\textheight]{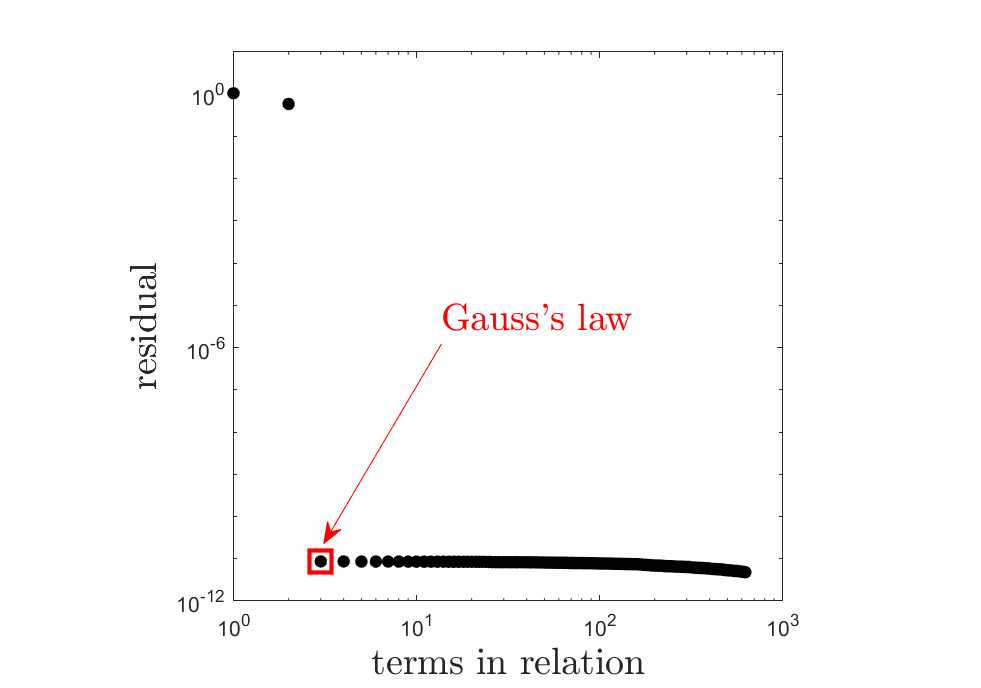}} \end{subfigure}
    \begin{subfigure}[]{\includegraphics[height=0.25\textheight]{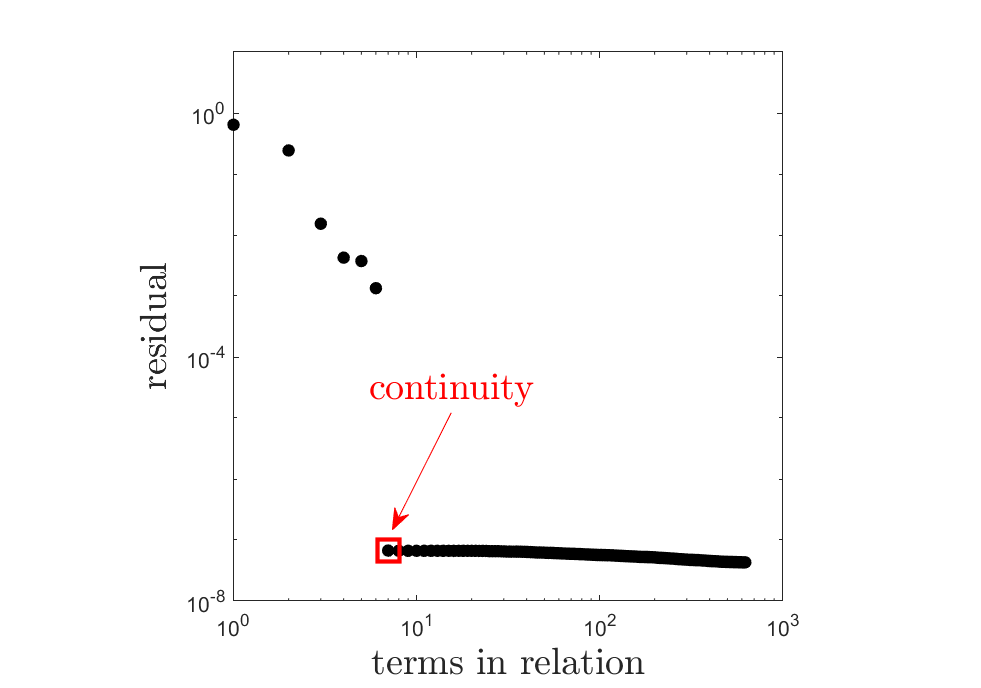}} \end{subfigure}
    \begin{subfigure}[] {\includegraphics[height=0.25\textheight]{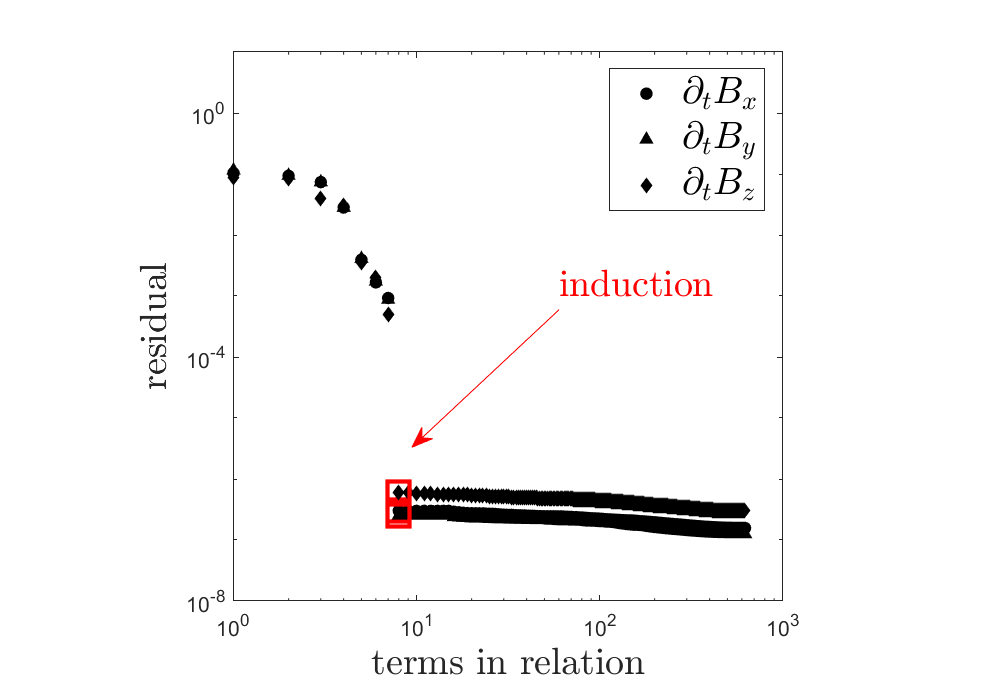}} \end{subfigure}
    \begin{subfigure}[] {\includegraphics[height=0.25\textheight]{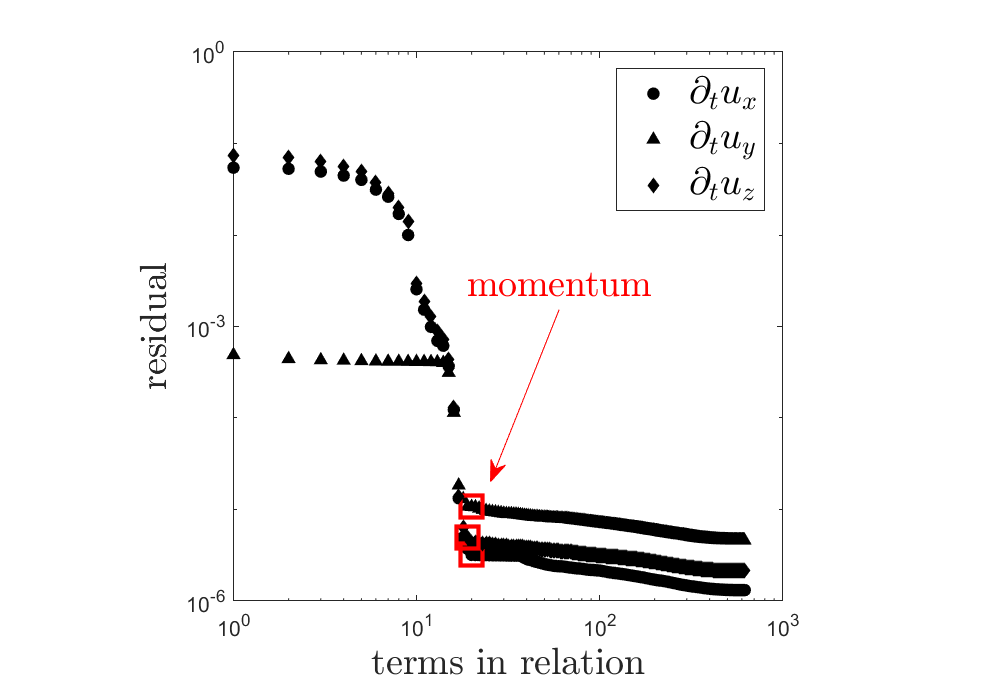}}\end{subfigure}

    \caption{Residual curves for all discovered models. (a) The identification of Gauss's law. (b) The identification of the continuity equation. The splitting of density into fluctuations makes the continuity equation a seven term model. (c) The three components of the induction equation. (d) The three components of the momentum equation.
    \label{fig:residuals}}
    \vspace*{-0.5cm}
\end{figure*}

This procedure generally finds models in order of increasing residual (see Fig.~\ref{fig:residuals}). The first model discovered is consistently Gauss's law, since this model is enforced exactly in the underlying simulation by evolving the vector potential ${\bf A}$. The continuity equation is found second with a residual $r\sim 10^{-7}.$ The components of the induction equation follow with residuals $r\sim 10^{-6.5}$. Lastly, the components of the momentum equation are identified with residuals of $r\sim 10^{-6}$. The $y$-component of the momentum equation is particularly interesting. The mean magnetic field $\langle {\bf B} \rangle = 0.1 \hat{y}$ makes the flow anisotropic. Removal of additional terms from the $y$-component of the momentum equation results in a residual curve distinct from the $x$ and $z$ components. This relation fails to indentify the dissipative term $\tilde{\rho} \partial_y^2 u_y$, since its magnitude is below the noise floor in the residual for that particular equation.

\bigskip

\noindent {\em Computational requirements.---\/} There are two computational intensive steps in solving the resulting $1376$ equations of the form~(\ref{eq:weak}) with each containing $650$ terms. The first step requires evaluating $1376\times 650=8.9\times 10^5$ weak-form integrals over $64^4=17\times 10^6$ grid points each or the equivalent of more than 10 TFLOP (tera floating point operations). Computing these integrals is a trivially parallelizable task: one can parallelize over distinct windows and over distinct library terms. Loading 3+1D data quickly becomes the bottleneck, so we parallelized the loading of all fields for 16 spacetime windows at a time. Then we evaluate all library term integrals for that data. Computing all 1376 integrals took $\sim 18$ hours on an Intel Xeon w9-3495X CPU with 56 cores.

The second step involves the greedy regression algorithm SPRINT. We implemented this with \texttt{MATLAB} on the same CPU. We parallelized across these cores the determination of the term to drop in each iteration, resulting in significant code acceleration. The 650 iterations required to identify each sparse equation were performed in approximately one minute. 

\bigskip

\noindent {\em Removing degeneracies among the learned equations.---\/} In order to discover the full set of possible equations, after one sparse equation is identified, we remove permanently from the library the term with maximal contribution $\| c_j {\bf g_j} \|$. We then restart the greedy procedure from the beginning, so that a new sparse equation can be identified. We halt this procedure when a closed model is recovered. 

It is important to emphasize here that using an implicit algorithm and incorporating a very large library of terms leads inadvertently our model discovery process to find a number of unexpected or redundant equations that may not deepen our understanding of the system. We discuss here two situations that lead to the discovery of additional equations.

If the dynamics of the system leads to correlations between some library terms, our use of an implicit algorithm discovers these emerging (non-dynamical) correlations as sparse models. For example, if the flow is weakly compressible, then volume integrals of different powers of the density fluctuation $(\rho - \rho_{\rm mean})^n$ will have a low residual and will be discovered. One would similarly expect emerging conservation laws to be also discovered in the same manner. We do not consider this to be a deficiency of the method: these correlations are as valid descriptions of the data as the governing equations and lead to important physical insights. This drove us to split the density as $\rho = \rho_{\rm mean} + \tilde{\rho}$ and greatly improved the success rate of sparse regression. 

The second situation arises by the fact that the product of any null sentence by any physical quantity is also a null sentence. For example, in addition to $\nabla \cdot {\bf B} = 0$, our algorithm also finds $\rho \nabla \cdot {\bf B} = 0$, $u_x \nabla \cdot {\bf B} = 0$, and many other algebraic consequences of the divergenceless magnetic field. Sparse regression is especially prone to finding $(\nabla \cdot {\bf B})^2 = 0$, since if $|\nabla \cdot {\bf B} | \ll 1$, then its square is certainly smaller. In our production run, this ended up not being a major issue. The linear model $\nabla \cdot {\bf B}$ was identified over its square. This is likely due to the accuracy of the weak formulation, whereas $(\nabla \cdot {\bf B})^2$ requires numerical differencing of the magnetic fields. We did establish a routine to check if a discovered equation could be factored as $(a + b + \cdots)(c+d+\cdots) = 0$ to within some numerical threshold. If the relation can be factored, the residuals of each factor are computed. If a factor had a sufficiently low residual, we would switch to that factor as the discovered model.

\bigskip

\noindent {\em Explicit forms of discovered equations.---\/}We present here the explicit forms of the discovered equations with their coefficients and the resulting residuals. We normalize the first term in the equation to a unity coefficient.

\begin{widetext}

\newcommand{\R}{\tilde{\rho}}
\begin{eqnarray}
&&\partial_xB_x + 0.999999999995 \partial_yB_y + 0.999999999999 \partial_zB_z = 0.0000000000087
\end{eqnarray}

\begin{eqnarray}
&&\partial_t\R + 0.9999999 \partial_xu_x + 0.9999999 \partial_yu_y + 0.9999999 \partial_zu_z + 1.0000005 \partial_x(u_x\R)\nonumber \\
&& + 1.0000005 \partial_y(u_y\R) + 1.00000002 \partial_z(u_z\R) = 0.000000067
\end{eqnarray}

\begin{eqnarray}
&&\partial_tB_z-1.00000009 \partial_x(B_xu_z)-1.000000003 \partial_y(B_yu_z) + 1.0000003 \partial_x(B_zu_x) + 0.9999996 \partial_y(B_zu_y)\nonumber \\
&&-0.0004000005 \partial_x^2B_z-0.000399999 \partial_y^2B_z-0.0003999998 \partial_z^2B_z = 0.00000025
\end{eqnarray}

\begin{eqnarray}
&&\partial_tB_x + 0.9999998 \partial_y(B_xu_y) + 1.0000002 \partial_z(B_xu_z)-1.00000009 \partial_y(B_yu_x)-0.9999998 \partial_z(B_zu_x)\nonumber \\
&&-0.0003999992 \partial_x^2B_x-0.000400004 \partial_y^2B_x-0.000399997 \partial_z^2B_x = 0.00000030
\end{eqnarray}

\begin{eqnarray}
&&\partial_tB_y-1.0000003 \partial_x(B_xu_y) + 1.0000002 \partial_x(B_yu_x) + 1.0000003 \partial_z(B_yu_z)-0.9999998 \partial_z(B_zu_y)\nonumber \\
&&-0.0004000009 \partial_x^2B_y-0.000400004 \partial_y^2B_y-0.000400001 \partial_z^2B_y = 0.00000060
\end{eqnarray}

\begin{eqnarray}
&&\partial_x\R + 0.999996 \partial_tu_x + 1.00004 \partial_t(u_x\R) + 1.000004 \partial_x(u_xu_x) + 0.999994 \partial_y(u_yu_x)\nonumber \\
&& + 1.000002 \partial_z(u_zu_x)-0.500008 \partial_x(B_xB_x)-0.9999994 \partial_y(B_yB_x) + 0.50000010 \partial_x(B_yB_y)\nonumber \\
&&-0.99998 \partial_z(B_zB_x) + 0.500010 \partial_x(B_zB_z)-0.0004002 \partial_x^2u_x-0.00039992 \partial_y^2u_x\nonumber \\
&&-0.0004001 \partial_z^2u_x + 1.00013 \partial_x (\R u_xu_x) + 0.9995 \partial_y (\R u_xu_y) + 1.00012 \partial_z (\R u_xu_z)\nonumber \\
&&-0.00034 \R \partial_x^2 u_x-0.00044 \R \partial_y^2  u_x-0.000386 \R \partial_z^2 u_x = 0.0000032
\end{eqnarray}

\begin{eqnarray}
&&\partial_z\R + 0.999998 \partial_tu_z + 1.00007 \partial_t(u_z\R) + 1.000003 \partial_x(u_zu_x) + 1.0000005 \partial_y(u_zu_y)\nonumber \\
&& + 1.000005 \partial_z(u_zu_z) + 0.500012 \partial_z(B_xB_x) + 0.500004 \partial_z(B_yB_y)-1.000004 \partial_x(B_zB_x)\nonumber \\
&&-1.000002 \partial_y(B_zB_y)-0.500002 \partial_z(B_zB_z)-0.00039996 \partial_x^2u_z-0.00040003 \partial_y^2u_z\nonumber \\
&&-0.0004008 \partial_z^2u_z + 0.99993 \partial_x (\R u_zu_x) + 1.0013 \partial_y (\R u_zu_y) + 0.9998 \partial_z (\R u_zu_z)\nonumber \\
&&-0.0003998 \R \partial_x^2 u_z-0.00044 \R \partial_y^2  u_z-0.00031 \R \partial_z^2 u_z = 0.0000044
\end{eqnarray}

\begin{eqnarray}
&&\partial_y\R + 0.99998 \partial_tu_y + 1.0002 \partial_t(u_y\R) + 0.999997 \partial_x(u_yu_x) + 0.99998 \partial_y(u_yu_y)\nonumber \\
&& + 1.000011 \partial_z(u_zu_y) + 0.49998 \partial_y(B_xB_x)-0.99998 \partial_x(B_yB_x)-0.49997 \partial_y(B_yB_y)\nonumber \\
&&-0.99997 \partial_z(B_zB_y) + 0.49998 \partial_y(B_zB_z)-0.0004002 \partial_x^2u_y-0.0004007 \partial_y^2u_y\nonumber \\
&&-0.0004003 \partial_z^2u_y + 1.003 \partial_x (\R u_yu_x) + 1.0012 \partial_y (\R u_yu_y) + 1.004 \partial_z (\R u_yu_z)\nonumber \\
&&-0.00041 \R \partial_x^2 u_y -0.000409 \R \partial_z^2 u_y = 0.000011
\end{eqnarray}

\end{widetext}


\end{document}